\documentclass[squareforqed]{llncs}

\usepackage{llncsdoc}
\usepackage{lipsum, graphicx}
\usepackage{stmaryrd}
\usepackage{verbatim}
\usepackage{answers}
\usepackage{xspace}
\usepackage{appendix}
\usepackage{listings}
\usepackage{wrapfig}
\usepackage{latexsym, amssymb,amsfonts,amstext,amsmath}
\usepackage{caption}
\usepackage{subcaption}
\usepackage{url}
\usepackage[colorlinks=true,
            urlcolor=blue,
            citecolor=blue,
            linkcolor=blue
            ]{hyperref}
\usepackage{sidecap}
\usepackage{wrapfig}
\newcommand{\mypara}[1]{\vspace{5pt}\noindent\textbf{#1.}}
\newcommand{\mysec}[1]{\section{#1}}
\newcommand{\mir}{\textrm{MIR}\xspace}
\newcommand{\mirs}{{\mir}s\xspace}

\newcommand{\figref}[1]{Fig.\,\ref{#1}}

\def\euler {\textsc{Euler}}
\def\cleantax {\textsc{CleanTax}}
\def\eulerx{\textsc{Euler}/X}
\def\eulerfo {\textsc{Euler}/FO}
\def\eulerasp {\textsc{Euler}/ASP}
\def\eulerpyrcc {\textsc{Euler}/PyRCC}

\title{\eulerx: A Toolkit for Logic-based Taxonomy Integration}
\titlerunning{Euler}

\author{Mingmin Chen\inst{1} \and Shizhuo Yu\inst{1} \and Nico
  Franz\inst{2} \and \\ Shawn Bowers\inst{3} \and Bertram
  Lud{\"a}scher\inst{1}}

  \institute{
Dept.\ of Computer Science, UC Davis, \email{\{michen,szyu,ludaesch\}@ucdavis.edu}
    \and
School of Life Sciences,    Arizona State University,  \email{nico.franz@asu.edu}
    \and
Dept.\ of Computer Science,    Gonzaga University,  \email{bowers@gonzaga.edu}
  }
\authorrunning{Mingmin Chen}

\begin{document}
\Opensolutionfile{appendix}

\maketitle

\begin{abstract}
  % We introduce the \eulerx\ Toolkit for logic-based taxonomy
  % integration in this paper.
  We introduce \eulerx, a toolkit for logic-based taxonomy
  integration.  Given two taxonomies and a set of alignment
  constraints between them, \eulerx\ provides tools for detecting,
  explaining, and reconciling inconsistencies; finding all possible
  merges between (consistent) taxonomies; and visualizing merge
  results. \eulerx\ employs a number of different underlying reasoning
  systems, including first-order reasoners (Prover9 and Mace4), answer
  set programming (DLV and Potassco), and RCC reasoners (PyRCC8). We
  demonstrate the features of \eulerx\ and provide experimental
  results showing its feasibility on various synthetic and real-world
  examples.
% The toolkit can make use of first order reasoners
%   (Prover9 and Mace4), answer set programming (DLV and Potassco), and
%   an RCC reasoner (PyRCC8). The Toolkit features alignment
%   inconsistency detection, inconsistency provenance, merging
%   taxonomies by generating all possible merges, and visualization of
%   the possible worlds. Here we demonstrate the features of \eulerx\
%   based on an abstract example at a small scale, and provide benchmark
%   values for larger scale implementations.
\end{abstract}

%%%%%%%%%%%%%%%%%%%
\mysec{Introduction}
\normalsize 

%%%%%%%%%       Nico's version         %%%%%%

%Biological classifications, or taxonomies, organize
%organismal groups into a 
%hierarchical system, thus facilitating the transmission and
%integration of relevant information about such groups.
%The creation and linkages among \emph{names} for taxa
%are regulated by \emph{Codes}
%of nomenclature. However, it is widely recognized that neither names 
%nor nomenclatural relationships (such as synonymy) are in themselves 
%sufficiently granular 
%to integrate the taxonomic entities occurring in related, succeeding 
%classifications~\cite{franz2008use,boyle2013taxonomic}. The \emph{taxonomic
%concept} approach~\cite{berendsohn1995concept} - i.e., the explicit handling of names and their circumscriptions in the context of particular classifications (e.g., \emph{Andropogon virginicus} L. [name] sec. Weakley, 2005 [classification]) \--- is suited to
%overcome these insufficiencies. Taxonomic Concepts (TCs) pertaining to
%multiple classifications ($T_1$, $T_2$) can be represented as separate 
%entities and linked via articulations ($A$), thereby allowing 
%reasoning approaches to perform the tasks of multi-taxonomy reconciliation and merge.

%%%%%%%%%       Shawn's version         %%%%%%

Biological taxonomies are hierarchical representations
used to specify formal classifications of organismal groups (e.g.,
species, genera, families, etc.)
While the names used for organismal groups (i.e., \emph{taxa}) are
regulated by various \emph{Codes} of nomenclature, it is widely recognized
that names alone are not sufficiently granular to integrate taxonomic entities
occuring in related
classifications~\cite{Kennedy05,franz2008use,boyle2013taxonomic}. Thus
additional information is required to relate taxonomic entities across
taxonomies. These relationships can then be used to compare different
taxonomies and integrate multiple taxonomies into a single
hierarchical representation.

% \normalsize Biological classifications (taxonomies) organize
% organismal groups (species, genera, families, etc.) into a
% hierarchical system, thus facilitating the transmission and
% integration of relevant information about such groups.
% The creation and linkages among \emph{names} for organismal groups
% (i.e., \emph{taxa}) are regulated by various codes of
% nomenclature. However, it is widely recognized that neither names
% nor nomenclatural relationships such as synonymy are in themselves
% sufficiently granular to integrate taxa occurring in related
% classifications~\cite{franz2008use,boyle2013taxonomic}.
% The use of \emph{taxonomic concepts}~\cite{berendsohn1995concept} are
% better suited for integration, since they allow for the explicit
% handling of names and their circumscriptions in the context of
% particular classifications (e.g., \emph{Andropogon virginicus}
% L. [name] sec. Weakley, 2005 [classification]) \--- is suited to
% overcome these insufficiencies. Taxonomic Concepts (TCs) pertaining to
% multiple classifications ($T_1$, $T_2$) can be represented as separate
% entities and linked via articulations ($A$), thereby allowing
% knowledge representation and reasoning approaches to perform the tasks
% of multi-taxonomy reconciliation and merge.
%

The first attempts to provide formal reasoning over taxonomies were
made in the MoReTax project \cite{berendsohn2003}, 
% BL: I don't think Berendsohn introduce RCC-5 relations.
% BL: But what they introduced turned out to be RCC-5, right?
which introduced the use of RCC-5 relations \cite{Randell92aspatial}
for defining relationships (articulations) among taxonomic concepts.
RCC-5 provides five basic relations for defining \emph{congruence},
\emph{proper inclusion}, \emph{inverse proper inclusion},
\emph{overlap}, and \emph{exclusion} among pairs of sets or concepts. These
comparative relations are intuitive to taxonomic experts who assert
them and who may also express ambiguity in their assessment
among concept pairs by using disjunctions of articulations: when the
exact relation is unknown to the expert, she can choose disjunctions of the
basic five relations, giving rise to up to 31 articulations, to capture partial knowledge.
For example, \emph{A \{congruence, overlap\} B} means the set $A$ can be
equivalent to or overlaps the set $B$.
The MoReTax approach was %significantly advanced and
formalized in first-order logic and implemented in \cleantax\
\cite{thau09}. This system implemented RCC-5 reasoning using the
first-order theorem provers Mace4 and Prover9 \cite{p9m4},
but also adding three taxonomic covering
assumptions---\emph{(i) non-emptiness}, \emph{(ii) sibling disjointness}, and
\emph{(iii) parent coverage}\footnote{Denoting that (i) concepts/taxa are non-empty,
i.e. have instances, (ii) sibling taxa are disjoint, (iii) the parent taxa is coverd by the union of child taxa, respectively.}---to achieve a working environment for
taxonomic reasoning.

Here we demonstrate the \eulerx\ toolkit which offers a suite of
interactive reasoning and visualization programs that extend the
capabilities of \cleantax\ while improving scalability. \eulerx\ also
adds new reasoning approaches to \cleantax\ including ASP (Answer Set
Programming \cite{answersets}) and a specialized RCC-8
reasoner~\cite{pyrcc8}. The toolkit implements a
comprehensive taxonomy import, merge, and visualization workflow,
with new features such as (1) PostgreSQL input of the original taxonomies
and expert-asserted articulations~\cite{franz2009perspectives}, (2)
detection of alignment inconsistencies, (3) diagnosis of inconsistency
provenance (based on provenance semirings~\cite{green2007provenance})
and interactive repair, (4) alignment ambiguity reduction, and (5)
visualization of merged taxonomies based on a set of inferred,
\emph{maximally informative relationships} (\mir) that reflect (6) one
or multiple possible worlds scenarios for taxonomy integration. We
illustrate these features using an abstract example that embodies
various of the aforementioned challenges (inconsistency, ambiguity,
multiple possible worlds) while maintaining close resemblance with
real-life use cases~\cite{franz2008use,Franz2013}.

\paragraph{Contributions.} \eulerx\ encodes the input taxonomies, articulations, and constraints
and feeds various inference problems to different reasoners (the
``X'' in \eulerx), then translates the output from those reasoners
to ``knowledge products'' to suit user needs.
%Jointly the reasoners and the wrapping are what we called \eulerx\ Toolkit. 
The main technical contribution are the ASP and other logical encodings, the
use of provenance, and result visualization, applied to real-world
taxonomy integration problems. 
To the best of our knowledge, \euler\ is the first system to apply
formal reasoning using ASP to such problems.

% This paper is aimed at a taxonomically and computationally literate
% audience wishing to utilize the Toolkit for this purpose. We note,
% however, that the \eulerx\ functions have wider relevance to the
% challenge of integrating nested datasets or ontologies in other
% non-biological domains.

% \paragraph{Outline.} The rest of this paper is organized as
% follows. Section~\ref{sec:sysdemo} contains
% preliminaries. Section~\ref{sec:exp} shows the experimental results
% and comparisons. Section~\ref{sec:con} concludes this paper and
% shows future work.

%%%%%%%%%%%%%%%%%%%
\mysec{System Demonstration}
\label{sec:sysdemo}

\begin{figure}[!h]
\centering
\includegraphics[width=3.6in]{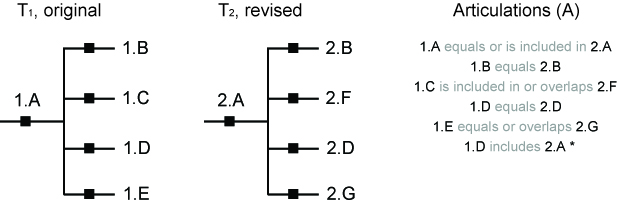}
\caption{\small Abstract example with two succeeding taxonomic
  classifications $T_1, T_2$ and a set of expert-asserted 
  articulations ($A$) among taxonomic concepts. 
Three articulations are disjunctive;  one (`*') leads to an inconsistency. $T_2$ (revised)
  builds on $T_1$ (original) but is a modification of $T_1$; it reuses
  $T_1$ entities but views and arranges them differently.}
\label{abs4}
\end{figure}

\mypara{Example} To demonstrate \eulerx, we introduce a simple example
(\figref{abs4}) of two taxonomies $T_1$ (original) and $T_2$
(revised). Each taxonomy includes only two levels (genus and species)
and ten constituent taxonomic concepts ($1.A$, $1.B$, $\ldots$, $2.A$, 
$2.B$, $\ldots$). Moreover there are six initial, expert-asserted
articulations that connect the respective entities. Three of these
include disjunctions (`or'), reflecting the expert's uncertainty as to
the precise relationship among concept pairs, and one leads to an
inconsistency (though the expert is not yet aware of this
error). Comparable, real-life examples are provided in
\cite{Franz2013}.
% BL: OLD: one articulation is inconsistent.
% BL: NEW: ... leads to an inconsistency. 
% (The articulation itself is NOT inconsistent!)

\begin{figure}[t]
\centering
\includegraphics[width=5.5in,keepaspectratio]{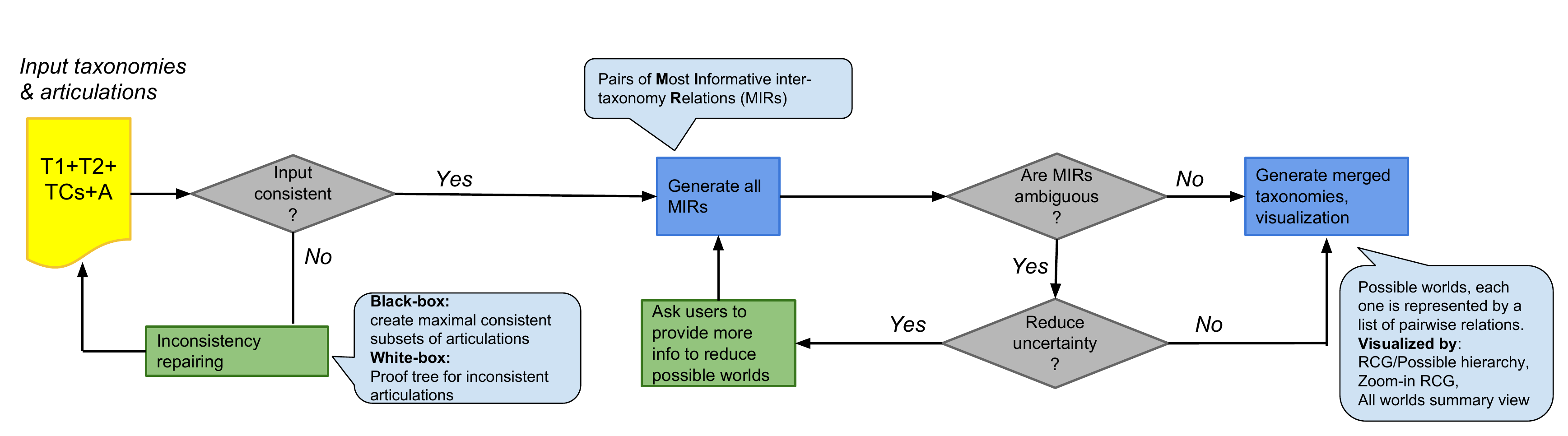}
\caption{\small \eulerx\ workflow overview: Input taxonomies
  $\mathsf{T}_1,\mathsf{T}_2$ together with expert articulations
  $\mathsf{A}$ and other taxonomic constraints $\mathsf{TCs}$ yield
  \mirs, merged taxonomies, and visualization products.}
% constraints  ,TCsyellow shape denotes
%   input; grey diamonds represent decisions; green
%   boxes show relevant processes; and blue boxes depict output
%   products.}
\label{eulerwf}
\end{figure}

\mypara{Workflow Overview} \eulerx\ will ingest the example input
(\figref{abs4}) into PostgreSQL in the form of three simple
spreadsheets: (1) a table that uniquely identifies each of the ten
taxonomic concepts; (2) a table that incorporates each set of five
concepts into its respective taxonomy ($T_1$, $T_2$) via \emph{is\_a}
parent/child relationships (e.g., $1.B$ $is\_a$ $1.A$, etc.); and
(3) a table with the six input articulations ($A$). The user also
specifies a set of taxonomic constraints (TCs), e.g.,
\emph{coverage}. The system then guides the user through an
interactive workflow (\figref{eulerwf}) that includes the following
major functions: consistency checking (including inconsistency
explanation and repair), \mir generation, ambiguity representation
(possible worlds\footnote{In each possible world, the relation of any
  two taxa is one of the RCC5.}) and reduction, and lastly output of
the merged taxonomies, including visualization and explanation of
the newly inferred \mirs. Jointly, these functions enable the expert to
obtain and comprehend a maximally consistent and unambiguous tabular
and graphic representation of the merged taxonomy. Alternative
reasoners---Prover9/Mace4 (FOL), DLV, Potassco (ASP), and PyRCC8
(RCC)---are integrated into the workflow to address specific reasoning
challenges.

%\begin{figure}[!htbp]

\begin{figure}[t]
\centering
\includegraphics[width=5in]{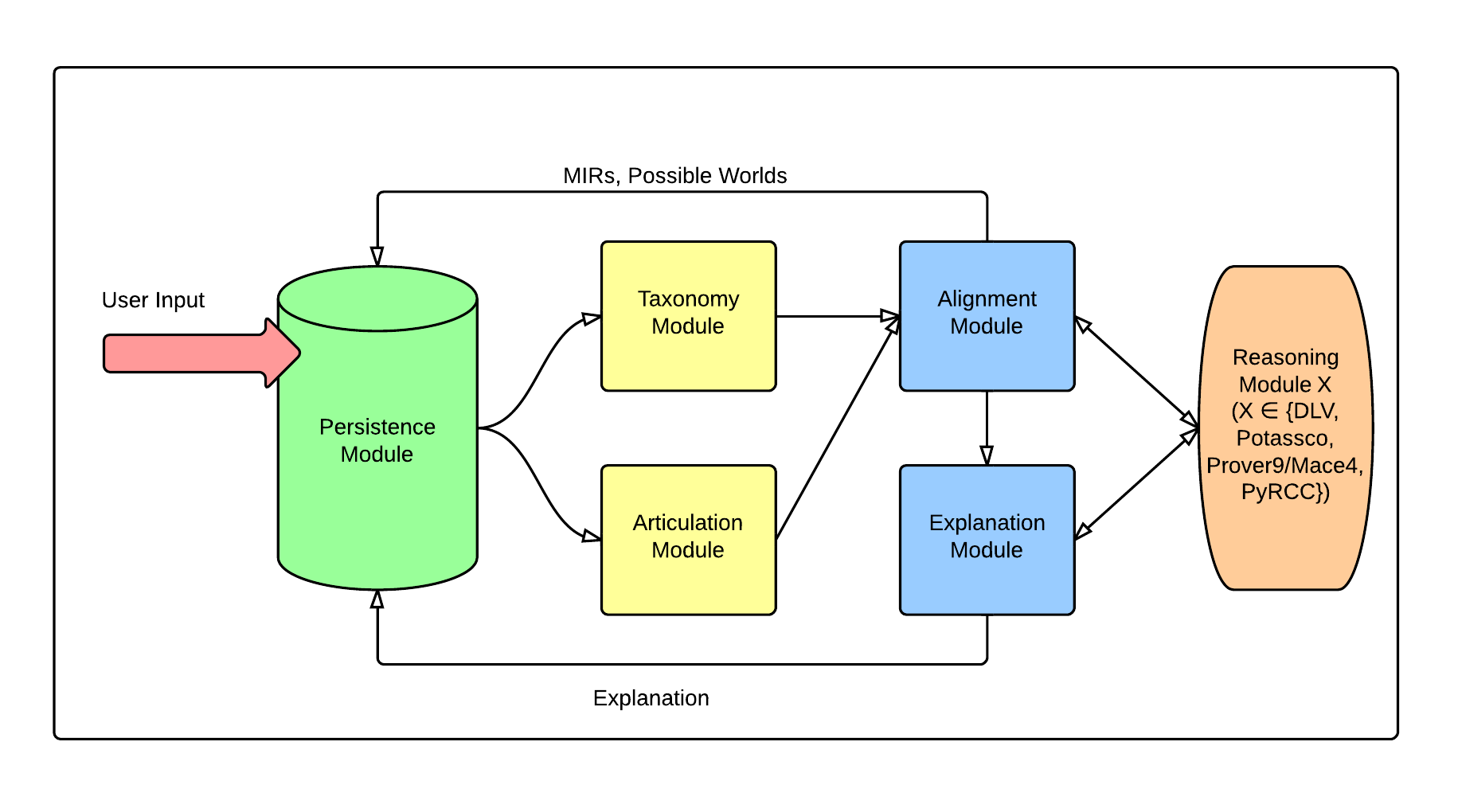}
\caption{\small \eulerx\ Toolkit Architecture.}
\label{archi}
\end{figure}

\mypara{Architecture} As shown in \figref{archi}, the \eulerx\ toolkit wraps six
modules: persistence module, taxonomy module, articulation module, alignment
module, explanation module, and reasoning module. User input will be stored in
the database (persistence module) after pre-processing; the taxonomy module and
articulation module load taxonomy and articulation data from the database, and
pass to alignment module; alignment module then generates inputs for the reasoning
module and determines the consistency and generate the possible worlds using
the results from the reasoning module. In case there is inconsistency, explanation
module will generate the provenance for the inconsistency based on the outputs
from reasoning module. The \mirs, possible worlds, and explanation will then be
passed to persistence module for storage. Reasoning module composes alternative reasoners,
such as Prover9/Mace4 (FOL), DLV \cite{citrigno1997dlv}, Potassco \cite{gebser2011potassco} (ASP), and PyRCC8
(RCC).

\mypara{Consistency Checking and Inconsistency Repair} The example
(\figref{abs4}) is computable in \eulerx\ using either FOL or ASP
reasoners (\figref{eulerwf}). The first processing step focuses on
testing the consistency of the input alignment (A). In our use case,
\eulerasp\ and \eulerfo\ both infer that
the input is inconsistent. In particular, \eulerfo\ provides a
black-box explanation that ``$1.D$ includes $2.A$'' is inconsistent
with the remaining articulations, and recommends removing this
articulation to obtain a consistent alignment. In contrast, \eulerasp\
offers a white-box explanation, stating that ``$1..D$ includes
$2.A$'' (implying that $1.D$ is a high-level, inclusive taxonomic concept) is
inconsistent with ``$1.A$ equals or is included in $2.A$'' and
``$1.D$ $is\_a$ $1.A$'' (jointly asserting that $1.D$ is a
low-level, non-inclusive concept). Thus one can repair the
inconsistency simply by deleting the articulation ``$1.D$ includes
$2.A$''. Based on subsequent \eulerx\ reasoning (\mir), we will
find that the correct $1.D$/$2.A$ articulation is ``$1.D$ is
included in $2.A$''.

\begin{figure}[!t]
\centering
\includegraphics[scale=0.22]{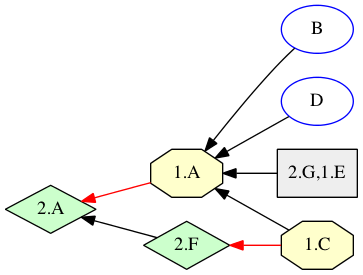}
\includegraphics[scale=0.22]{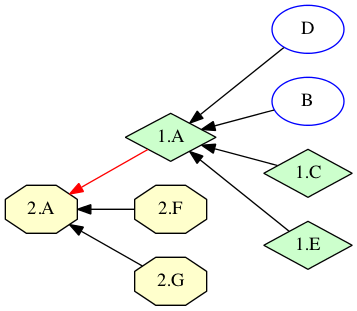}
\includegraphics[scale=0.22]{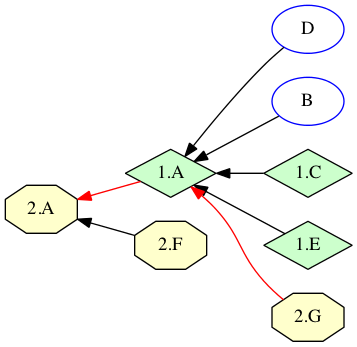}
\includegraphics[scale=0.22]{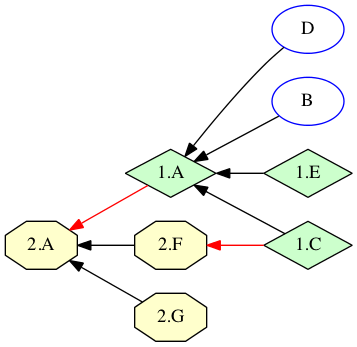}
\\
\includegraphics[scale=0.22]{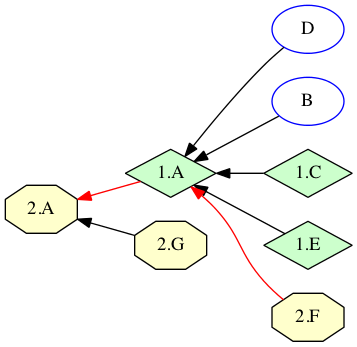}
\qquad\qquad
\includegraphics[scale=0.22]{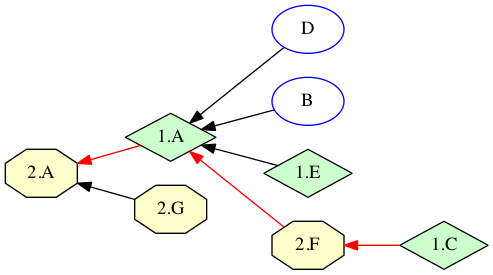}
\qquad\qquad
\includegraphics[scale=0.25]{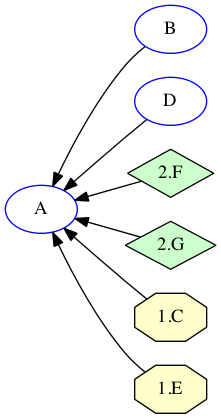}
\caption{\small Set of possible worlds $w_0, \dots, w_6$ (i.e.,
  concept hierarchies or RCGs) resulting from the \mir inferred by
  \eulerx\ based on the repaired input example: white ellipse (grey box) nodes show
  congruent (merged) concepts in both taxonomies; green diamond/yellow octagon nodes show
  concepts unique to each taxonomy; black edges show input $is\_a$
  relations; and red edges show newly inferred $is\_a$ relations.
  RCGs do not represent concept overlap.}
\label{7pws}
\end{figure}

\mypara{Generating \mir and Possible World Visualizations} Once the
input example's inconsistency is repaired, \eulerx\ will proceed to
generate all \emph{maximally informative relations} (\mir; see Thau
\emph{et al.}, 2009~\cite{thau09}) among taxonomic concept pairs. The
interaction of the three articulations involving disjunction (\figref{abs4})
form an inherently ambiguous input alignment, which results in a total
of seven equally consistent ``possible world'' solutions. These
possible worlds can be displayed using a simple ``Reduced Containment
Graph'' -- a transitively reduced directed graph
in which an edge represents proper inclusion.  (\figref{7pws}).
%\footnote{These graphs cannot adequately represent taxa overlap.}

%\begin{figure}[!t]
%\centering
 % \includegraphics[scale=0.4]{abstract4_cl_neato.pdf}%
  %\caption{\small Visualization of possible worlds for the input
    %example, where the distance between two possible worlds is the
    %shortest distance traceable in the graph (e.g., the distance
    %between worlds 5 and 6 is 4).}
  %\label{pwcluster}
%\end{figure}

\begin{wrapfigure}{r}{0.5\textwidth}
\vspace{-20pt}
  \centering
  \includegraphics[scale=0.5]{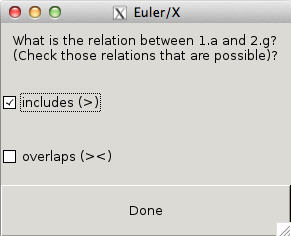}
  \captionof{figure}{Pop-out interactive window showing ambiguity reduction.}
  \label{ar}
\vspace{-30pt}
\end{wrapfigure}
\mypara{Interactive Ambiguity Reduction} Although the seven possible
worlds (\figref{7pws}) accurately reflect the resolving power of the
input alignment (\figref{abs4}), the user may now have the ability and
desire to reduce the inherent ambiguity by selectively eliminating
certain (apparently improbable) possible worlds. This is facilitated
by the \eulerx\ feature of ambiguity reduction. At runtime, \euler\
asks the user more questions (generated by a decision tree
function) via pop-out interactive windows allowing the user to
select the preferred answer, e.g., by specifying that the current
articulation in the query instance is ``\texttt{1.A $>$ 2.G}'', i.e.,
\texttt{1.A} properly includes \texttt{2.G} (\figref{ar}). Based on
the responses \eulerx\ can reduce the number of possible worlds from
seven to three, filtering out four possible worlds in which
\texttt{1.A} and \texttt{2.G} and overlap.

\begin{wrapfigure}{r}{0.5\textwidth}
\vspace{-30pt}
  \centering
  \includegraphics[scale=0.4]{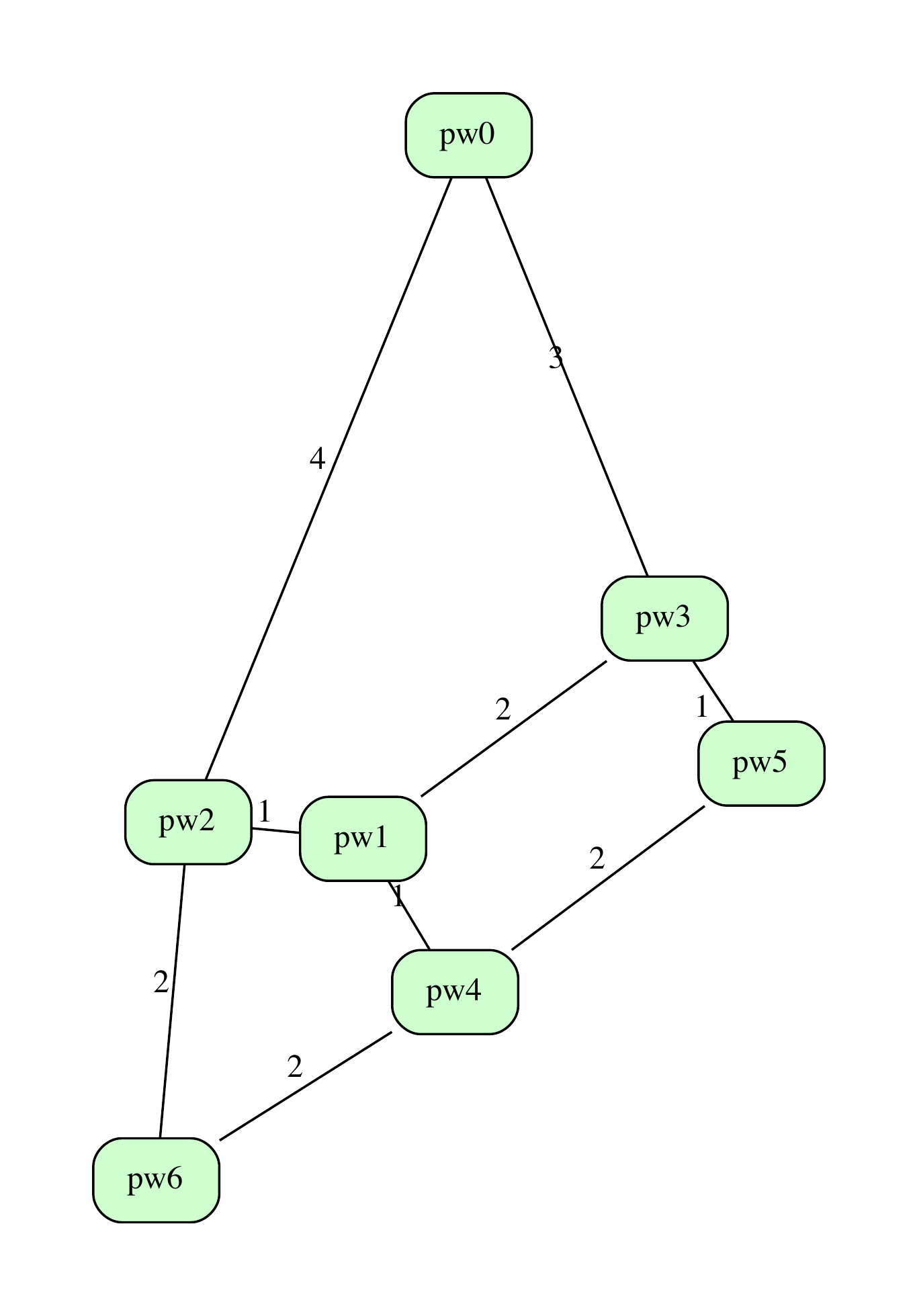}
\vspace{-20pt}
  \captionof{figure}{Distance matrix-based visualization of the seven possible worlds of the input example. See also \figref{7pws}. The absolute distance between two possible worlds is the shortest distance traceable in the graph; e.g. the distance between possible worlds 5 and 6 is 4 steps.}
  \label{pwcluster}
\vspace{-30pt}
\end{wrapfigure}

\mypara{Visual clustering of similar possible worlds} We can expect
some use cases with larger sized input taxonomies and multiple
inherent ambiguities to yield large numbers of possible
worlds. \eulerx\ offers a visual representation of the cumulative
possible worlds ``universe'' via a distance matrix
(\figref{pwcluster}). As shown in \figref{7pws}, our input example has
seven possible worlds. We can compute pairwise distances among these
by integrating the numbers of \mirs in which they differ and thereby
generating a network that summarizes the similarities and differences.

\mypara{Additional features} \eulerx\ also provides information on the
\emph{provenance} of a newly generated \mir relation. Moreover the
toolkit can provide users with a consensus perspective of all possible
worlds, i.e., specifying what is true in all of them, or how often a
particular \mir occurs across all possible worlds.

%%%%%%%%%%%%%%%%%%%
\mysec{Performance Results}
\label{sec:exp}

We tested the performance and scalability of different reasoning
approaches, including \eulerfo\ (Prover9/Mace4), \eulerasp\ (DLV and
Potassco), and \eulerpyrcc\ (PyRCC8). Tests used both real-life and
simulated examples as well as performed both consistency checks and
 \mir and possible worlds computation. The running time was measured
using increasingly larger input datasets.  All examples were tested on
an 8-core, 32GB-memory Linux server.

%\TODO{For \eulerfo\ this needs to be explained better. To generate mirs, it makes $m*n$ calls (in the worst case)?} 
While \eulerfo\ checks consistency by calling Mace4 once and then
generates each \mir by calling Prover9\footnote{To get a \mir, Prover9
  is called to answer ``yes'' or ``no'' to the five base relation
  questions.}  (for $m*n$ \mir's assuming there are $m,n$ entities in
each taxonomy), the other \euler\ tools only invoke the reasoner once

%\begin{wrapfigure}{r}{0.6\textwidth}
\begin{wrapfigure}{r}{0.65\textwidth}
%\vspace{-40pt}
\centering
\includegraphics[width=3in]{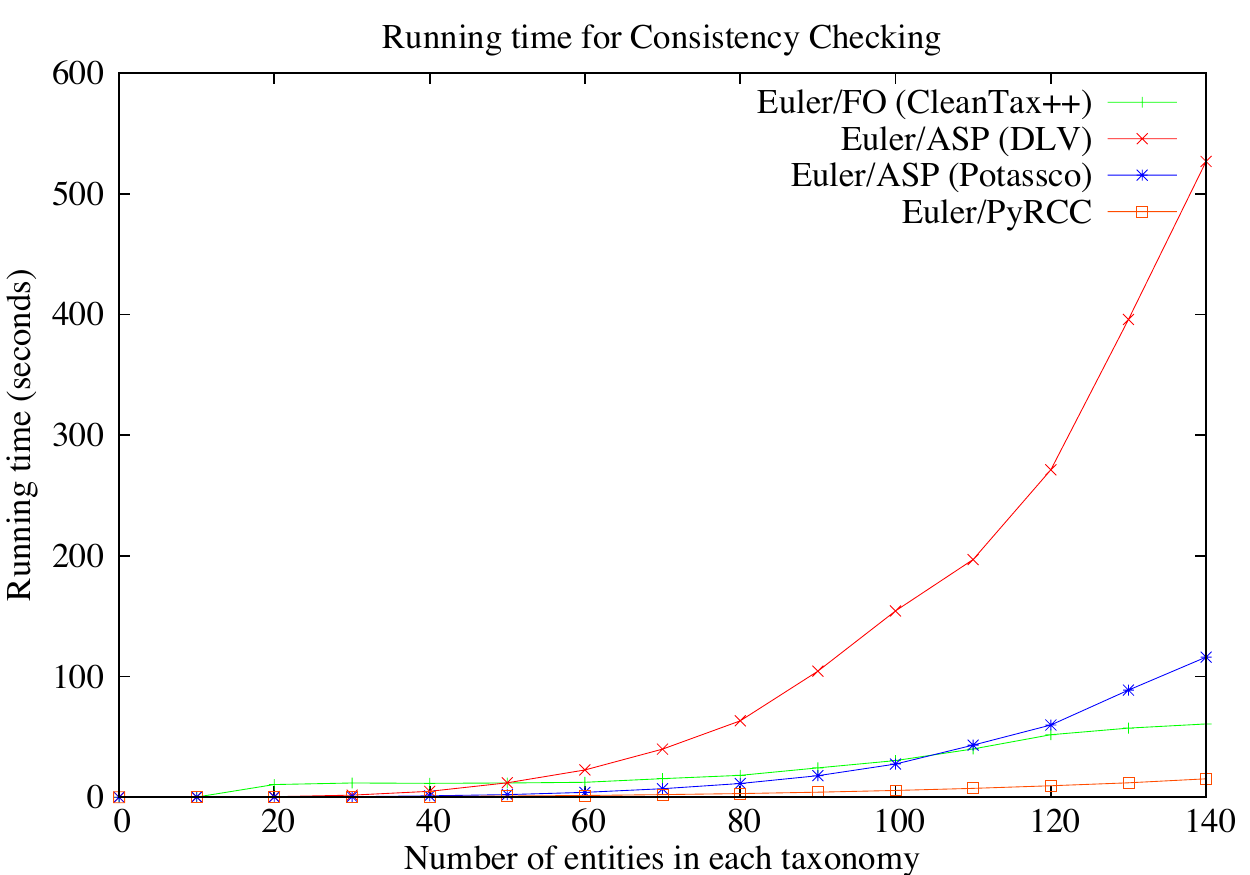}
\includegraphics[width=3in]{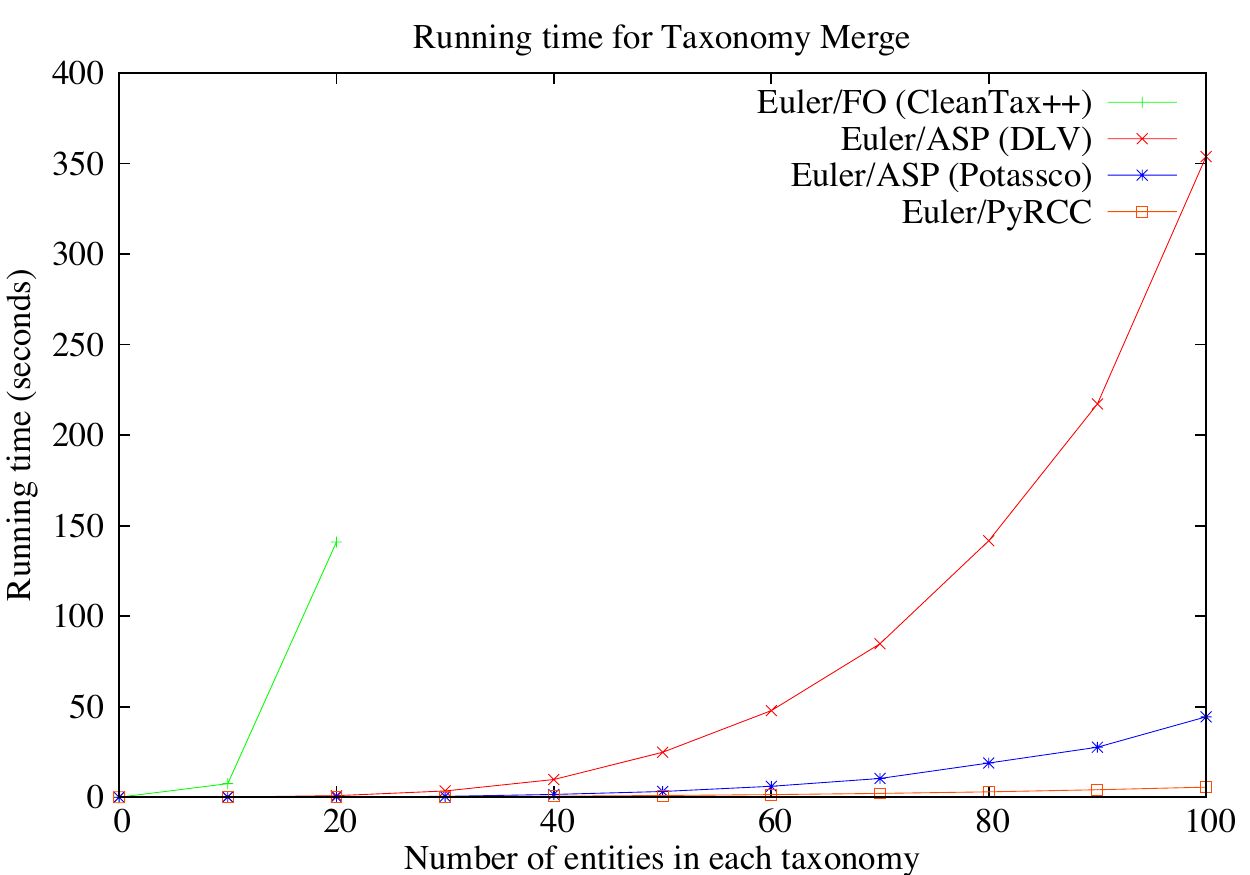}
\caption{\small Running times for consistency checking (left) and taxonomy
  merge (right) on  synthetic taxonomies (balanced taxonomy trees of depth 8
  with ``is included in'' articulations, resulting in a single possible world).}
\label{exp_merge1}
\vspace{-10pt}
\end{wrapfigure}

\noindent to check consistency and merge taxonomies (\mir and possible world
generation). This is why \eulerfo\ is faster for consistency checking
(specifically, \eulerfo\ is slower than \eulerasp\ (Potassco) when the
number of entities in each taxonomy is less than 100, but faster when
it is more than 100), but very slow in \mir generation as shown in
\figref{exp_merge1}. For taxonomy merge, PyRCC8 is faster than
Potassco, Potassco is faster than DLV, and DLV is much faster than our
FO-based approach. However, note that \eulerpyrcc\ is not capable of
applying the same merge as the other tools since the coverage
constraints cannot be asserted using RCC-5.  When considering all
three taxonomic constraints, the Potassco-based \euler\ is the fastest
and reasonably good overall, since it can perform taxonomy merge for
realistic taxonomies of 100 entities in half a minute.

%%%%%%%%%%%%%%%%%%%
\mysec{Conclusions and Future Directions}
\label{sec:con}

\eulerx\ is open source and can be downloaded from
BitBucket\footnote{\href{https://bitbucket.org/eulerx/euler-project}{https://bitbucket.org/eulerx/euler-project}}. Planned
future developments include: (1) support for incremental changes to
alignments; % . At present the \eulerx\ Toolkit implementation rechecks
% the entire alignment whenever it checks incrementally, as users add,
% modify or remove articulations in the alignments, or modify the
% taxonomies.
(2) an improved ASP-based tool, using the results from PyRCC8; (3)
development of a user-friendly GUI; and (4) further exploration of 
other reasoners, e.g., those developed for OWL.

\mypara{Acknowledgements}
We thank the anonymous reviewers for their helpful comments.
Work supported in part by NSF awards IIS-1118088 and DBI-1147273.
%\vspace{-5pt}

%
%This toolkit provides various features for merging taxonomies: it can detect the inconsistency in the input alignment; it provides provenance-based explanation for inconsistency and new knowledge; it generates / visualizes the merged taxonomies; it persists input/output data in Postgres such that users may query immediately or in the future easily etc. Among all the \eulerx\ tools, Potassco-based tool is the fastest one for taxonomy merge with consideration of coverage taxonomic constraint. However, we still have several improvements and features as the future work.
%

%\newpage
%%%%%%%%%%%%%%%%%%%
\bibliographystyle{abbrv}
%\scriptsize
\bibliography{eulerdemo-final}
\normalsize

\appendix

\end{document}